\documentclass[12pt,preprint]{aastex}

\shorttitle{Turning the Tides on ComBer and UMa\,II}
\shortauthors{Mu\~noz et al.}

\begin{document}

\title{Turning the Tides on the Ultra-Faint Dwarf Spheroidal Galaxies: Coma Berenices and 
Ursa Major II\altaffilmark{1}}

\author{
Ricardo R. Mu\~noz\altaffilmark{2},
Marla Geha\altaffilmark{2}, \&
Beth Willman\altaffilmark{3}}

\altaffiltext{1}{Based on observations obtained at the Canada-France-Hawaii Telescope (CFHT) which is 
operated by the National Research Council of Canada, the Institut National des Sciences de l'Univers 
of the Centre National de la Recherche Scientifique of France,  and the University of Hawaii.}

\altaffiltext{2}{Astronomy Department, Yale University, New Haven, CT 06520, USA 
(ricardo.munoz@yale.edu, marla.geha@yale.edu)}

\altaffiltext{3}{Haverford College, Department of Astronomy,
370 Lancaster Avenue, Haverford, PA 19041, USA (bwillman@haverford.edu)}

\begin{abstract}

We present deep CFHT/MegaCam photometry of the ultra-faint Milky Way satellite
galaxies Coma Berenices (ComBer) and Ursa Major II (UMa\,II).
These data extend to $r\sim25$, corresponding to three magnitudes below the main 
sequence turn-offs in these galaxies.
We robustly calculate a total luminosity of $M_{V}=-3.8\pm0.6$ for ComBer and
$M_{V}=-3.9\pm0.5$ for UMa\,II, in agreement with previous results. 
ComBer shows a fairly regular morphology with no signs of active tidal
stripping down to a surface brightness limit of $32.4$ mag arcsec$^{-2}$. Using a
maximum likelihood analysis, we calculate the half-light radius of ComBer
to be $r_{\rm half}=74\pm4$\,pc ($5.8\pm0.3\arcmin$) and its ellipticity $\epsilon=0.36\pm0.04$.  
In contrast, UMa\,II shows signs of on-going disruption.  We map its morphology down
to $\mu_{V}=32.6$ mag arcsec$^{-2}$ and found that UMa\,II is larger than previously
determined, extending at least $\sim700$\,pc ($1.2^\circ$ on the sky) and it is also
quite elongated with an ellipticity of $\epsilon=0.50\pm0.2$.
However, our estimate for the half-light radius, $123\pm3$\,pc ($14.1\pm0.3\arcmin$)
is similar to previous results. We discuss the implications of these findings 
in the context
of potential indirect dark matter detections and galaxy formation.
We conclude that while ComBer appears to be a stable dwarf galaxy, UMa\,II shows
signs of on-going tidal interaction.

\end{abstract}


\section{Introduction}

For decades, only about a dozen dwarf galaxies were known to orbit the 
Milky Way.  The majority of these systems corresponded to dwarf spheroidal 
(dSph) galaxies, the least luminous, but, by number, the dominant galaxy type 
in the present-day universe.
However, over the last five years, and thanks to the advent of the Sloan Digital Sky 
Survey (SDSS; \citealt{York2000}) the field of dwarf galaxies in the Milky Way has 
been revolutionized.  To date, fourteen new systems have been detected as slight 
overdensities in star count maps using the SDSS data (\citealt{Willman2005a,Willman2005b}; 
\citealt{Belokurov2006,Belokurov2007a,Belokurov2008,Belokurov2009}; 
\citealt{Zucker2006a,Zucker2006b}; \citealt{Sakamoto2006}; \citealt{Irwin2007}; 
\citealt{Walsh2009}).
These recent discoveries have revealed a previously unknown population of ``ultra-faint'' 
systems which have extreme low luminosities, in some cases as low as $L_{V}\sim300$\,L$_{\sun}$ 
\citep{Martin2008}, and in average comparable (or lower) to 
those of Galactic globular clusters. However, spectroscopic surveys of the
majority of these systems reveal kinematics and metallicities in line
with those of dwarf galaxies (\citealt{Kleyna2005}; \citealt{Munoz2006}; 
\citealt{SimonGeha2007}; \citealt{Kirby2008}; \citealt{Koch2009}; \citealt{Geha2009}).

Dynamical mass estimates of the ultra-faint galaxies based on line-of-sight radial 
velocities indicate that these galaxies are extremely dark matter dominated, with central
mass-to-light ratios (M/L) as high as $1,000$ in solar units.  These systems are 
thus good laboratories to constrain cosmological models \citep[e.g~the 'missing 
satellites problem';][]{Kauffmann1993,Moore1999,Klypin1999,SimonGeha2007} and to study the 
properties of dark matter (\citealt{Strigari2008}; \citealt{Kuhlen2008}; \citealt{Geha2009}).  
However, these applications hinge critically on the assumption that the masses and 
density distributions in these systems are accurately  known.  Current mass estimates 
for the ultra-faint dwarfs are based on the assumption that the dynamical state of these 
systems have not been significantly affected by Galactic tides, and therefore that they are 
near dynamical equilibrium.

There is circumstantial 
evidence for past tidal disturbance in a number of these satellites based on 
 morphological studies.  \citet{Coleman2007} for instance, found the Hercules 
dSph to be highly elongated, with a major to minor axis ratio of 6:1, larger than for 
any of the ``classical'' dSphs, and argued for a tidal origin of this elongation.  
\citet{Belokurov2007a} reported fairly distorted morphologies for several of the new 
ultra-faint dwarfs, 
although \citet{Martin2008} 
showed that
results based on low number of star may not be statistically significant 
because they suffer from shot noise.
The strongest evidence for tidal interaction perhaps is 
for Ursa Major II, which appears to be broken into several clumps, lies close to the 
great circle that includes the Orphan Stream (\citealt{Zucker2006b}; \citealt{Fellhauer2007}) 
and shows a velocity gradient along its major axis \citep{SimonGeha2007}.

If dSphs galaxies are currently undergoing tidal stripping, or have in the past, then 
kinematical samples are expected to be ``contaminated'' with unbound or marginally bound 
stars. This will impact subsequent dynamical modeling, often resulting in overestimated 
mass contents (\citealt{Klimentowski2007}; \citealt{Lokas2009}).  The degree of such 
contamination is a function of both the projection of the orbit along the line-of-sight 
and the strength of the tidal interaction.  Therefore, even though tides do not affect 
appreciably the inner kinematics of dSphs until the latest stages of tidal disruption 
(e.g., \citealt{OLA95}; \citealt{Johnston1999}; \citealt{Munoz2008}; \citealt{Penarrubia2008}), 
studies aimed at identifying the 
presence of tidal debris can help elucidate the dynamical state of these dwarf galaxies.  
Obvious tidal features around dSphs would indicate the presence of unbound stars and their 
effects on the derived masses would need to be investigated.  Alternatively, a lack of 
clear detections in the plane of the sky would narrow the possibilities that kinematical 
samples suffer from contamination, although it would not automatically imply that an object 
has not been tidally affected. Tidal features could still exist but be aligned preferentially 
along the line-of-sight and thus be hard to detect.

Ultra-faint dwarf galaxies are useful probes of galaxy formation on the smallest scales 
(\citealt{Madau2008}; \citealt{Ricotti2008}).  One of the outstanding questions related to 
their discovery is whether these galaxies formed intrinsically with such low luminosities, or 
whether they were born as brighter objects and attained their current luminosities through 
tidal mass loss. Current metallicity measurements (e.g., \citealt{Kirby2008}) support the former
scenario. They show that the ultra-faint dwarfs are also the most metal poor of the dSphs,
following the luminosity-metallicity trend found for the classical dSphs\footnote{\citet{SimonGeha2007},
using the same spectroscopic data but a different technique, reported systematically higher 
metallicites for the ultra-faint dwarfs than those of \citet{Kirby2008}. The discrepancy is
explained by the fact that the former study used the \citet{Rutledge1997} method
that relies on the linear relationship between Ca II triplet equivalent widths and [Fe/H], a 
technique now known to fail at very low metallicities.}.  
On the other hand, if tidal features around these objects are firmly detected, it would clearly 
support the latter hypothesis.  

Given the importance of the questions at hand, it is essential 
that we investigate the dynamical state of these satellites.
Due to the extreme low luminosity of the ultra-faint dwarfs, and therefore
low number of brighter stars available for spectroscopic studies, deep imaging 
is currently the only way to efficiently detect the presence of faint morphological 
features in the outskirts of these systems.  
We expect any tidal debris to be of very low surface brightness (\citealt{Bullock2005}) 
so that detecting it via integrated light is virtually impossible. However, these galaxies 
are sufficiently nearby to resolve individual stars. Thus, using matched-filter techniques 
it is possible to detect arbitrarily low surface brightness features ($\sim 35$ mag arcsec$^{-2}$,  
\citealt{Rockosi2002}; \citealt{Grillmair2006}). 

In this article we present the results of a deep, wide-field photometric survey of 
the Coma Berenices (ComBer) and Ursa Major II (UMa\,II) dSphs,
claimed to be two of the most dark matter dominated dSphs \citep{Strigari2008},
carried out with the MegaCam imager on the 
Canada-France-Hawaii Telescope. 
In \S2, we present details about the observations and data reduction, as well as
of artificial star tests carried out to determine our completeness levels and photometric
uncertainties. In \S3, we recalculate the structural parameters of these systems using
a maximum likelihood analysis similar to that of \citet{Martin2008}. In \S4 we 
present the results of our morphological study. We discuss the statistical significance
of our density contour maps and the robustness of our result to the variation of 
different parameters. We show that ComBer looks fairly regular in shape and find 
no signs of tidal debris down to a surface brightness limit of 32.4 mag 
arcsec$^{-2}$, whereas UMa\,II shows a significantly elongated and distorted shape, 
likely the result of tidal interaction with the Milky Way. Our discussion and conclusions
are presented in \S5 and \S6, respectively.

\section{Observations and Data Reduction}

\subsection{Observations}

Observations of both ComBer and UMa\,II were made with the MegaCam imager 
on the Canada-France-Hawaii Telescope (CFHT) in queue mode. 
MegaCam is a wide-field imager consisting of $36$ $2048\times4612$ pixel CCDs,
covering almost a full $1\times1$ deg$^{2}$ field of view with a pixel
scale of $0.187\arcsec$/pixel.
UMa\,II was observed on nights between February and March of 2008, whereas
ComBer was observed between April and May of 2008.

For each dSph, four different, slightly overlapping fields were observed for a 
total area coverage of nearly $2\times2$ deg$^{2}$ in the case of ComBer, and
$1.4\times2$ deg$^{2}$ in the case of UMa\,II. In each field, the center of the 
dSph was placed in one of the corners so that, when combined, the galaxy is located 
at the center of the overlapping regions as shown in Figure 1.
For each of the four fields, we obtained eleven 270s dithered exposures in 
Sloan $g$ and eleven 468s dithered exposures in Sloan $r$ in mostly dark conditions 
with typical seeing of $0.7-0.9\arcsec$.
The dithering 
pattern was selected from the standard MegaCam operation options in order to cover 
both the small and large gaps between chips (the largest vertical gaps in MegaCam 
are six times wider than the small gaps).

\subsection{Data Reduction and Astrometry}

MegaCam data are pre-processed by the CFHT staff using the ``Elixir'' package 
(\citealt{Magnier2004}) prior to delivery. The goal is to provide the user with 
frames that are corrected for the instrumental signature across the whole 
mosaic. This pre-process includes bad pixel correction, bias subtraction and flat 
fielding. A preliminary astrometric and photometric solution is also included in 
the pre-processed  headers.

The World Coordinate System (WCS) information provided with the data is only
approximate, and we refine it using
the freely available
SCAMP\footnote{See http://astromatic.iap.fr/software/scamp/} package 
as follows: 
Terapix SExtrator\footnote{See http://astromatic.iap.fr/software/sextractor/} is run on 
all 
chips (SCAMP reads in the output files generated by SExtrator) 
and output files are written in the FITS-LDAC\footnote{LDAC stands for Leiden Data Analysis Center.} 
format. SCAMP is then run on all 
chips.
SCAMP uses the approximate WCS information in the frames' headers as a starting point, 
and then computes astrometric solutions using, in our case, the SDSS-Data Release 6 (DR6, 
\citealt{York2000}; \citealt{Adelman2008}) reference catalog which is automatically 
downloaded from the Vizier database.  Typically, several hundred stars in common between 
each of our chips and SDSS are used to compute
the astrometry and final solutions. In ComBer, fields have typical global astrometric 
uncertainties 
of $rms\sim0.15\arcsec$, while for fields in UMa\,II they are slightly higher, 
with $rms~\sim0.2\arcsec$. 
The output from SCAMP is a FITS header file (one per processed frame), which is then used 
to update the WCS information for that given chip.
The updated headers are then used to translate $x$ and $y$ positions into final
equatorial coordinates.

\subsection{Point Source Photometry}

Prior to performing point-source photometry on our data, we split each mosaic frame
into its 36 individual chips.
To take full advantage of our 11 dithered exposures per field (and per filter), we 
test two different methods for carrying out the photometry.

\subsubsection{ALLFRAME on Individual Exposures}

For the first method, we carry out our photometry running first DAOPHOT/Allstar
on the individual, non-coadded frames and then running the ALLFRAME package on the resulting files
as outlined in \citet{Stetson1994}.
ALLFRAME performs photometry simultaneously on all 22 frames for a given field (11 per filter).
DAOPHOT/Allstar are required prior to ALLFRAME to determine point spread function (PSF) solutions 
for each chip as well as to generate starlists for them individually. 
Optimum starlists to be used as input by ALLFRAME are obtained by cross matching the 
DAOPHOT/Allstar results for the individual frames using the DAOMATCH and DAOMASTER packages 
(\citealt{Stetson1993}). These packages also provide reasonably good estimates of the offsets
between dithered individual exposures necessary to run ALLFRAME.
Final output files from ALLFRAME are then combined into a single catalog.

\subsubsection{DAOPHOT/ALLSTAR on Stacked Images}

In our alternative reduction method, we coadd all 11 exposures per filter for each of 
the $36$ individual chips using the SWARP package\footnote{See http://astromatic.iap.fr/software/swarp/}. 
SWARP uses the WCS information stored in the frames' headers to correct for shifts, small 
rotation between chips and distortions.  However, since original WCS information in the 
headers is only approximate, we refine the astrometry (as described in \S2.2) for all frames 
prior to combining.
SWARP then coadds the frames and the DAOPHOT II/Allstar 
package is used to do point-source photometry on the stacked images.

Using artificial star tests described below, we determine that the ALLFRAME method (\S2.3.1) yields 
slightly deeper and more accurate photometry and therefore we continue our data reduction 
and analysis using the catalog generated as described in the first part of this section.

\subsubsection{Photometric Calibration}

Finally, photometric calibration was carried out by comparison with the SDSS-Data Release 7 
(DR7, \citealt{Abazajian2009}) catalog.  We first match our final photometry with the SDSS 
stellar catalog. We typically found several hundred stars in common with SDSS per chip. To 
determine zero point and color terms we only use stars with $18<r<21.5$ and $0.0<g-r<1.0$. 
The brighter limit is given by the saturation limit of our CFHT data.
We then fit the equations:

\begin{eqnarray}
g=g_{\rm instr} + g_{0} + g_{1}(g-r)
\end{eqnarray}

\begin{eqnarray}
r=r_{\rm instr} + r_{0} + r_{1}(g-r)
\end{eqnarray}

\noindent where $g_{\rm 0}$ and $r_{\rm 0}$ are the zero points and $g_{\rm 1}$ and $r_{\rm 1}$ are 
the respective color terms.
We do this for each chip individually in order to determine whether there
are chip-to-chip variations. In all cases, we find that the chip-to-chip differences, for both
zero points and color terms are lower than the uncertainties in the derived parameters, 
and therefore we combine the stars in all 36 chips to derive final zero point and color term
values. We do this via a linear least-squares fit weighting by the respective uncertainties
in the photometric magnitudes (as estimated by ALLFRAME) and rejecting $3\sigma$ outliers.
We obtain zero point and color terms for each mosaic field independently (four per dSph, 
eight total). The resulting constants differ by less than 2\% in all cases.
Uncertainties in the zero points vary from $0.003 - 0.004$\,mag whereas uncertainties
in the color terms are of the order of $\sim0.005$.

\subsection{Artificial Star Tests}

Final photometric uncertainties and completeness levels are determined via artificial star
tests. 
We first generate a fake color-magnitude-diagram (CMD) from which we will randomly select
artificial stars. We populate the CMD with stars in the magnitude and color ranges of
$18<g/r<28$ and $0.0<(g-r)<1.0$ respectively, with four times more stars in the $23<g/r<28$
range than in the brighter half of the CMD.
We then select one of the 11 dithered exposures as our reference frame. In all cases we
select the first exposure in the $g$ filter for this purpose. The goal is to inject
stars in all 22 frames (per chip) in the same fake RA and DEC positions in order to mimic
real observations. 
Artificial stars are then randomly selected from the fake CMD and are injected into the 
reference frame in a uniform grid with a spacing of $40$\,pixels in both the $x$ and $y$ directions. 
Using the refined astrometric solution calculated by SCAMP we convert the $x$ and $y$ positions 
of the artificial stars into RA and DEC coordinates which are then converted back to $x$ and $y$ 
positions, but in the reference frame of the other exposures. Care is taken so that all the stars
fall in the common area between the 22 different exposures. In the end, nearly $3,950$ stars are
introduced per chip. We repeat this procedure ten times to improve our statistics, each time 
randomly offsetting the grid's zero-point position in $x$ and $y$. We then carry out photometry 
on the artificial stars using the ALLFRAME method in the exact same manner as we did for the 
science frames.

We perform this test on only one chip per mosaic frame due to computational constraints, for a total 
of four chips in each dSph. Since each of these fields was observed under slightly different seeing 
and darkness conditions, their respective magnitude limits are also slightly different. Thus, to 
set conservative overall completeness levels we select the values from our ``shallowest'' field in 
each dSph as our final numbers.
We find that this represents a better choice than determining our limits from our combined 
artificial star's photometry for all four chips. In the latter case, the deepest fields dominate 
the results.
The 50\% and 90\% overall completeness levels of our photometry thus correspond to 
$g=25.8$ and $25.2$ and $r=25.4$ and $24.75$, respectively, for our ComBer fields, 
and $g=26.0$ and $25.4$ and $r=25.5$ and $24.9$, respectively, for UMa\,II.
Figure 2 shows the resulting completeness levels.

In order to clean our catalogs of galaxy interlopers and other, non-stellar detections, 
we apply cuts using DAOPHOT's sharpness ($sharp$) and $\chi$ parameters.
To define appropiate cuts, we fit third-degree polynomials to the $\chi$ and $sharp$ 
distributions, as functions of $g$ magnitude, obtained from the artificial star tests.
After applying these cuts to both our photometric catalog and artificial star photometry
our 90\% completeness levels drop to $g=24.8$ and $r=24.4$ for ComBer and $g=24.9$ and $r=24.5$ for 
UMa\,II. 

In Figures 3 and 4, we show color-magnitude-diagrams (CMD) for the central regions of both
galaxies, including 90\% completeness levels. In both cases, we reach
at least three magnitudes below the main sequence turn-off of these systems,
improving by roughly an order of magnitude, with respect to the original SDSS photometry,
the number of likely members of each galaxy.
Photometric uncertainties as a function of magnitude were derived by taking the difference 
between the artificial star's actual and measured magnitude
and are shown as error bars in Figures 3 and 4.

\section{Results}

\subsection{Structural Parameters}

Our photometric catalogs for both ComBer and UMa\,II contain an order of magnitude
more stars than the SDSS photometry, and thus provide us with an excellent 
opportunity to re-estimate the structural properties of these systems.
\citet{Martin2008} carried out a comprehensive analysis of the structural parameters
of all the ultra-faints using the original SDSS photometry. However, in the cases of ComBer
and UMa\,II, owing to their very low luminosities ($M_{V}\sim-4$), only about $100$ 
and $300$ stars were found to belong to the respective galaxy 
and therefore the derived parameters (and morphologies) suffer from significant uncertainties
due to low numbers statistics.

For each galaxy, we re-calculate the photometric center ($\alpha_0$, $\delta_0$), 
ellipticity ($\epsilon$), position angle ($\theta$), half-light radius\footnote{In the
case of King profiles we calculate their $r_{\rm core}$ and $r_{\rm tidal}$.}
($r_{\rm half}$) and background density ($\Sigma_{b}$). 
We try three different density profiles, exponential,
Plummer (\citealt{Plummer1911}) and empirical King (\citealt{King1962}), and obtain structural 
parameters for all three:

\begin{equation}
\Sigma_{exp}(r)=\Sigma_{0,E}{\rm exp}\left({-{r \over r_{E}}}\right)
\end{equation}

\begin{equation}
\Sigma_{\rm Plummer}(r)=\Sigma_{0,\rm P}\left(1+{r^{2} \over r_{P}^{2}}\right)^{-2}
\end{equation}

\begin{equation}
\Sigma_{\rm King}(r)=\Sigma_{0,\rm K} \left(\left(1+ \frac{r^{2}}{r_{c}^{2}}\right)^{-\frac{1}{2}} - \left(1+ \frac{r_{t}^{2}}{r_{c}^{2}}\right)^{-\frac{1}{2}}\right)^{2}
\end{equation}

\noindent where $r_{E}$ and $r_{P}$ are the exponential and Plummer scale lengths and $r_{c}$ and $r_{t}$ correspond
to the King core and tidal radii respectively. The exponential scale length is related to the
half-light radius by the relation $r_{\rm half}=1.68\times r_{E}$ while in the case of the
Plummer profile, $r_{P}$ is equivalent to $r_{\rm half}$. 

We have followed a procedure similar to the one outlined in \citet{Martin2008} which relies on 
a maximum likelihood (ML) analysis of the data to constrain the structural parameters. 
The basic idea of the method is as follows:
We assume that the positions of the stars are well represented by a given density profile
(one of the three mentioned above) which is, in turn, well described by a set of parameters
$p_{1}$,$p_{2}$,...,$p_{j}$. We then maximize a function of the form:

\begin{equation}
L(p_{1},p_{2},...,p_{j}) =\prod_{i}l_{i}(p_{1},p_{2},...,p_{j})
\end{equation}

\noindent where $l_{i}(p_{1},p_{2},...,p_{j})$ is the probability of finding the datum $i$ 
given the set of parameters $p_{1}$, $p_{2}$,...,$p_{j}$.
In the case of an exponential profile, this function takes the form:

\begin{equation}
l_{i}(p_{1},p_{2},...,p_{j})=S_{0}{\rm exp}\left(-\frac{r_{i}}{r_{E}}\right)+\Sigma_{b}
\end{equation}

\noindent where $S_{0}$, $r_{i}$ and $r_{E}$ are expressed in terms of the structural
parameters we want to determine.

In practice, we look for a global maximum $L(\hat{p_{1}}, \hat{p_{2}},...,\hat{p_{j}})$ by searching 
the $j$-dimensional parameter space. In our case, the parameter space is $6^{th}$-dimensional, 
with the free parameters $\alpha_{0}$, $\delta_{0}$, $\epsilon$, $\theta$, $r_{half}$ and $\Sigma_{b}$.
In the case of a King profile, we fix the
background density using the value obtained for the Plummer profile because of a degeneracy
between $\Sigma_{b}$ and the King tidal radius.
To find a solution, we use the method described in \citet{Sand2009} which relies on the amoeba simplex 
algorithm (\citealt{Press1988}) to search the parameter space. 
This method is somewhat sensitive to the specified region of parameter 
space to be searched (i.e. the initial guess and allowed range for the parameters) but it runs considerably 
faster than using an iteratively refined grid.
In order to derive uncertainties for the structural parameters, we carry out $10,000$ 
bootstrap (resampling with replacement) realizations of our data. 
The distribution of a given parameter is well described by a Gaussian with only minor 
deviations in some cases (King core and tidal radii in particular), and therefore we fit
Gaussian functions and report their mean and standard deviation as the mean and 1-$\sigma$ 
uncertainty for a given parameter.  

To facilitate comparison with previous studies, we first compare
the results of our algorithm when applied to the SDSS data with those derived by
\citet{Martin2008}.
To mimic their star selection procedure as closely as possible, we select stars 
within $1$ degree from the satellites' centers, impose the conditions that $r<22.0$, $g<22.5$ 
and, additionally, that stars live near the M92 fiducial line shifted to the distance 
of the respective dwarf satellite.  We perform the comparison only for the exponential profile.
For both galaxies, the resulting structural parameters are in very good agreement. 
However, since we determine our parameters via a bootstrap analysis, our derived uncertainties 
are significantly larger. 
\citet{Martin2008} did not perform a bootstrap analysis of their data and 
estimated their uncertainties using the statistical properties of the $L$ function. However,
as shown in the same study, statistical fluctuations due to the low number of stars in these
dSphs can dramatically affect the morphology as well as the best-fitting parameters and
therefore we regard our uncertainties as more realistic.

We then apply the ML algorithm to our dataset.
We select stars above the 90\% completeness levels and 
within a region around the main sequences of these objects (in order
to improve the signal to noise ratio of dSph versus Milky Way foreground stars). 
For this purpose, we use an isochrone for a population $13$\,Gyr old and [Fe/H]$=-2.27$ 
($Z=0.0001$) from \citet{Girardi2004} 
and define a region around the main sequences by the condition
that stars have to lie within $0.075$\,mag from the isochrone. 
We varied this distance between $0.050$--$0.100$\,mag and found 
no significant changes in the final parameters.
We note that we do not match the isochrone to the blue edge
of the main sequence as it is customary when isochrone matching is used to derive star
cluster properties, but instead place the isochrone so that it goes through the
middle of the main sequences.
For ComBer, our ML method yields a $r_{\rm half}=5.8\pm0.3$ arcmin, an ellipticity $\epsilon=0.36\pm0.04$ 
and a position angle $\theta=-67.0\pm3.6$
for the case of an exponential profile with the parameters derived for the other profiles being
very similar.
In the case of UMa\,II we find $r_{\rm half}=14.1\pm0.3$ arcmin, $\epsilon=0.50\pm0.02$ and $\theta=-74.8\pm1.7$.
The final sets of structural properties for both ComBer and UMa\,II are presented in Table 1.

This exercise also allows us to compare the results obtained from both our CFHT photometry and the 
shallower SDSS data. 
As it was pointed out by \citet{Sand2009} in their work on the Hercules dSph, 
we find relatively good agreement between the results obtained with both datasets, but 
our photometry allows us to place much tighter constrains
on the derived structural properties. This is not surprising given that 
our deeper photometric database is less sensitive to shot-noise effects that 
plague the shallower SDSS data.
Figure 5 illustrates this point. In this figure we show the distribution of four 
structural parameters for ComBer, $\alpha_{0}$, $\delta_{0}$, $r_{\rm half}$ and $\epsilon$ 
for all $10,000$ bootstraps and for both datasets where the smaller uncertainties 
derived with the CFHT data are evident.

In Figures 6 and 7a we show background subtracted density profiles for both satellites
with the best exponential, Plummer and King models overplotted. 
These are not fits to the binned data points, but are
models constructed with the best parameters found via the ML method.
As it can be seen, in the case of ComBer, both the exponential and King profiles are
adequate descriptions of the data, with the Plummer profile being perhaps a less adequate one. 
The case of UMa\,II is quite different. Neither profile does a good job matching the data. In fact,
as shown in Figure 7b, the data are better matched by a shallower inner power law ($\gamma=-0.96$) 
and a steeper one in the outer parts ($\gamma=-2.40$) reminiscent of the density profile
derived by \citet{Grillmair2009} for the Bo\"{o}tes III stellar overdensity or the inner
power law shown by the tidally stripped Palomar~5 (Pal~5) globular cluster \citep{Odenkirchen2003}.
This result 
is consistent with a scenario wherein UMa\,II has been significantly tidally stripped, as 
suggested by \citet{Zucker2006b}.
We will explore this possibility in more depth in \S4.

\subsection{Absolute Magnitude}

We estimate the absolute total magnitudes of ComBer and UMa\,II following a procedure similar
to that outlined by \citet{Walsh2008}. 
This method relies solely on the number of stars belonging to the galaxy, and not on 
their individual magnitudes.
As pointed out by \citet{Martin2008}, \citet{Walsh2008} and \citet{Sand2009}, the extreme 
low luminosity of the ultra-faint galaxies and therefore the low number of stars they 
contain, make traditional methods, like the addition of fluxes from individual stars, 
too sensitive to the inclusion (or exclusion) of potential members (outliers).
Adding or subtracting a few red-giant-branch stars (with absolute magnitudes as
bright as $M_{V}=-4.8$ for a metallicity of [Fe/H]$\sim-2.5$) can significantly alter the 
total luminosity measurement in these cases.
We therefore follow the method briefly described below.

We first assume that the stellar populations of both systems are mainly comprised by old, 
metal-poor stars, and therefore are well described by a single population. 
We can then model that population with a theoretical luminosity function to help estimate its
luminosity.   For this exercise, we adopt luminosity functions from \citet{Girardi2004} for 
a population $13$\,Gyr old and with $Z=0.0001$ (which corresponds to [Fe/H]$=-2.27$). 
This assumption is reasonable in light of the metallicity measurements by \citet{Kirby2008} 
who show that both dwarf galaxies have mean metallicities around [Fe/H]$=-2.5$.
In addition, we use theoretical luminosity functions computed using two different
initial mass functions (IMF): a \citet{Salpeter1955} with cutoff at $0.01$\,M$_{\sun}$ and 
a \citet{Chabrier2001} log-normal.

The theoretical luminosity function gives us the relative number of stars in magnitude
bins, which can be integrated to obtain the total flux down to a given magnitude limit.
One of the parameters we determine using the ML algorithm is the background surface density
$\Sigma_{b}$ which is related the number of stars $N_{*}$ that belong to the galaxies by:

\begin{equation}
N_{*}=N_{\rm total} - A\Sigma_{b}
\end{equation}

\noindent where $N_{\rm total}$ is the total number of stars used to derive the structural parameters and $A$
represents the total area of our fields.
We use this $N_{*}$ to normalize the theoretical luminosity function.
By integrating the luminosity function and correcting by this normalization factor we
obtain the actual flux corresponding to our galaxies down to the respective magnitude limits. 
To account for the light contributed by stars fainter than this limit, we
add the remaining normalized integrated flux.

As mentioned above, this method differs from traditional ones,
and our
derived luminosities are conceptually different from those derived for brighter galaxies. 
In our method, two different galaxies comprised of the same population and
with exactly the same number of stars will always have the same luminosity. However, in 
practice this will most likely not be the case as one might expect galaxies with similar 
populations to show an intrinsic spread in luminosities even if they have the same number 
of stars.  To account properly for this
we estimate uncertainties in our derived
absolute magnitudes by carrying out a bootstrap analysis.
The procedure is as follows: We treat the theoretical luminosity function used to calculate
the luminosities as a cumulative probability function (down to our 90\% completeness levels)  
of the number of stars expected as a function of magnitude. We then randomly draw a number $N_{*}$ 
of stars from the luminosity function and add their fluxes. We do this 10,000 times. 
Using this method, for the case of a Salpeter IMF we find, for ComBer, 
$M_{V}=-3.8\pm0.6$ (after using $V-r=0.16$
from \citealt{Girardi2004}). For UMa\,II we obtain $M_{V}=-3.9\pm0.5$. 
For the second choice of IMF we obtain $M_{V}=-3.8\pm0.6$ for ComBer and $M_{V}=-4.0\pm0.6$
for UMa\,II. The uncertainties quoted here do not include the uncertainties in the distance
to this objects.
To test further whether our method yields reliable values, we apply it to SDSS data selected
in the same way as described in \S3.1. For a choice of a Salpeter or log-normal IMF our results 
are 
almost identical to those of \citet{Martin2008} but as in the case of deriving structural
parameters our uncertainties are larger.

\section{Morphology}

A main goal of this study is to re-assess to what extent Galactic tides may be affecting 
the structure of ComBer and UMa\,II.  Ever since the early numerical simulations of disrupted 
satellites by \citet{PP95} and \citet{OLA95}, 
the observed elongation of dSph galaxies have been associated with the degree of 
their tidal interaction with the Milky Way\footnote{The same studies, and more recently \citet{Munoz2008}
have pointed out that this is not usually the case. Intrinsically spherical satellites are tidally elongated
only when the satellite has become nearly unbound.}.
In \S3.1 we found that ComBer is slightly elongated ($\epsilon=0.36$)
roughly along the direction of the Galactic center (shown by the solid line in Figure 8).  
UMa\,II, on the other hand, is much more elongated, with $\epsilon=0.5$ but its elongation 
is in the east-west direction, nearly perpendicular to the direction of the Galactic 
center\footnote{This does not preclude UMa\,II from being elongated along the direction
of the Galactic center, but such elongation would be mostly along the line-of-sight
and therefore very difficult to detect.}
(UMa\,II has galactic coordinates of [l,b]=[152.5, 37.5]).
While alignment with the Galactic center is expected in intrinsically elongated satellites, 
it is not a sufficient condition to infer active tidal stripping in the case of ComBer. 
Tides can still affect a system without necessarily stripping stars off of that 
system (C. Simpson \& K. Johnston, in preparation).

The quality of our photometric data, which reach at least three magnitudes below the 
main sequence turn-off in both galaxies, allows us to address the tidal stripping question 
more directly by studying the morphology of these satellites, this time much more reliably 
than in previous studies.  
To look for potential tidal features, we create smoothed isodensity contour maps. 
The photometric catalogs used to create these maps are the same as those we used
to derive structural parameters, and include all star-like objects (i.e., after
$sharp$ and $\chi$ cuts) down to our 90\% completeness levels that live
in a region around the main sequence as defined in \S3.1.
To make the maps, the positions of the stars are binned into $40\arcsec \times40\arcsec$ bins
which are subsequently spatially smoothed with an exponential filter of scale $2\arcmin$. 
Other reasonable bin sizes and exponential scales were tried as well, with no significant change
in the overall results.

\subsection{Coma Berenices}

Figure 8 shows the resulting map for ComBer.
This object shows fairly regular contours similar
to those found using SDSS photometry \citep{Belokurov2007a} but at a much higher 
statistical significance.
We do not detect 
signs of potential tidal 
debris down to the  $3\sigma$ isodensity contour 
level (the lowest one shown in Fig. 11) which corresponds to a surface brightness of 
$\sim32.4$ mag arcsec$^{-2}$.
The lack of significant elongation, clumpiness or irregularities in the morphology of ComBer
indicate that it is unlikely that it is currently being significantly affected by tides.
This, coupled with the kinematics of ComBer measured by \citet{SimonGeha2007} and the
corresponding mass derivation, supports the interpretation that ComBer is in fact a stable
dwarf galaxy. 
Even if ComBer is being tidally perturbed at levels below our detection limits, 
such low-level effects are not likely to alter its inner kinematics appreciably
(e.g., \citealt{Read2006}; \citealt{Munoz2008}; \citealt{Penarrubia2008}).

We note that our countour maps do not show the actual smoothed surface 
brightness contours, but instead show ``significance contours'', i.e, for each dwarf, 
we calculate the mean surface density and its standard deviation in areas away from 
the region dominated by the galaxy (background density), and 
plot the background-subtracted local density divided by the standard deviation.
If the values of pixels in our surface density map are well described by a Gaussian distribution,
then the standard deviation used here would measure precisely how significant our features are
in terms of $\sigma$ values. However, the distribution of pixels will, most likely, not be
well described by a Gaussian, and therefore we need an experiment to determine the true
significance of features in our data.
To do this we first randomize the position of the 
stars, but leave the photometry untouched. We then select stars photometrically and generate 
smoothed density maps in the same way as we do for the actual data.
When we perform this test, we find that $3\sigma$ overdensities randomly scattered across the
field are not uncommon, but higher significance features are very rare. We therefore conclude 
that isolated $3\sigma$ features detected in our real maps are very likely random noise and 
do not reflect potential tidal features.

To assess both the robustness of the overall shape of ComBer and the significance of apparent
substructure in its outer density contours, specifically a hint of elongation in the 
northeast-southwest direction, we carry out a different test. In this case, we bootstrap our 
photometric samples (as opposed to randomly assign positions to the stars) and redo the maps.  
Figure 9 shows eight different bootstrap realizations of the data.
It can be inferred from
the figure that 
hints of substructure in the outer parts of ComBer are not statistically
significant, but the overall shape and structure of this system are well
established with our data. This is not surprising, given the improvement
in the number of stars that belong to the satellite achieved with our photometry.

We conclude that ComBer is unlikely to be significantly affected by Galactic tides
and therefore it represents a solid case of a stable dwarf galaxy whose characteristic
size is smaller than $\sim120$\,pc. 

\subsection{Ursa Major II}

In contrast to the regular morphology of ComBer, UMa\,II looks entirely different. 
Figure 10 shows its isodensity contour map,
where the $3\sigma$ contours are equivalent to $32.6$ mag arcsec$^{-2}$.
Our photometry confirms previous findings that UMa\,II is highly elongated and it 
shows that UMa\,II is larger than previously reported, extending at least three times 
beyond its measured half-light radius, or nearly $700$\,pc ($\sim1.2^\circ$) on the sky.
More striking perhaps, is the fact that contours of UMa\,II looks more like a boxy-like system 
than an elliptical one. Only the very inner parts of UMa\,II resemble a 
spheroidal object.  

In Figure 10 we also show the best-fit orbit derived for UMa\,II by \citet{Fellhauer2007}, 
based on the assumption that it is the progenitor of the ``Orphan Stream'' (\citealt{Zucker2006b}; 
\citealt{Belokurov2007b}). This shows that the observed east-west elongation of UMa\,II does 
not match the predicted direction from the model, although this is not enough to rule out a 
connection between the two systems.

To assess the statistical significance of our results we carry out the same tests 
we described in section \S4.1. 
We find that remaking the surface density maps after randomly assigning coordinates to 
the stars yields identical results as in the ComBer case, namely, that isolated 
$3\sigma$ features are likely background noise but higher significance ones are real.
Likewise, making contour maps after bootstrapping the data shows that the overall shape of 
UMa\,II is fairly well established and is insensitive to resampling.
We illustrate this in Figure 11 where we show isodensity contours for eight 
different bootstrap realizations of our UMa\,II data. 

\citet{Zucker2006b} found tentative evidence that UMa\,II might be broken into several 
clumps (see their Figure 1). Our contour map for this object does not show statistically 
significant substructure. However, we are able to reproduce this results if we make contour maps 
use only stars brighter than $g=23$.
We regard the presence of substructure in the inner parts of UMa\,II as statistical fluctuations 
in its density due to low number of stars.

Another of the ultra-faints that looks similarly elongated as UMa\,II is the Hercules dSph. 
A recent study of this object by \citet{Sand2009}, who use deep photometry obtained
with the Large-Binocular-Telescope (LBT), show that Hercules has an ellipticity $\epsilon=0.67$
and extends at least $500$\,pc ($\sim13\arcmin$ on the sky). We measured an ellipticity $\epsilon=0.5$ for UMa\,I, but 
while this object resembles Hercules in this regard, UMa\,II's morphology is even more irregular.

\section{Discussion}

 Our understanding of the ultra-faint Milky Way satellites hinge on
having reliable morphologies and robust structural parameters.
Tidal disruption can result in disturbed morphologies and tidal
features which can be very low surface brightness, yet have a profound
affect on the interpretation of a given object.   Deep photometry is
the only means to assess the presence or absence of faint tidal
structure.   In the absence of complete disruption, well determined
sizes and luminosities are critical in calculating the total mass of a
system.  Our main goal in this paper is to search for signs of tidal
features and determine structural parameters around the Milky Way
satellites ComBer and UMa\,II.   Our $r-$ and $g$-band CFHT photometry
reaches three magnitudes below the main sequence turn-off in these
systems, corresponding to stars which are two to three magnitudes
fainter than currently accessible to spectroscopic studies.

We achieve similar surface brightness limits for both ComBer and
UMa\,II ($\sim 32.5$ mag arcsec$^{-2}$), yet find very different
morphologies for these two ultra-faint satellites.  As seen in
Figures~8 and 9, ComBer is remarkably regular and devoid of potential
tidal features, in stark contrast to the elongated and irregular
structure seen in UMa\,II.  Since low number of stars can produce
spurious features and/or shapes that can later be interpreted as signs
of tidal stripping \citep{Martin2008}, we have outlined in \S4.1
two tests which demonstrated the robustness of our maps.  We show that
while the $3\sigma$ contours in these figures lack statistical
significance, all higher $\sigma$ contours are robust to resampling
and should be regarded as a solid result.

ComBer is remarkably regular and devoid of morphological features
potentially due to tidal debris
down to a surface brightness level of $32.4$ mag arcsec$^{-2}$.  We
have found a slight elongation in the direction toward the Galactic
center but no discernible irregularities are visible.   We cannot of
course completely rule out the presence of unbound debris at fainter
surface brightnesses, but, if present, it should reflect only mild,
low-level tidal effects.  Similarly, we cannot rule out the presence
of tidal tails along the line-of-sight.  In the kinematic survey of
ComBer, \citet{SimonGeha2007} report a velocity dispersion
for this object of $4.6\pm0.8$\,km\,s$^{-1}$\ with only one possible
interloper star beyond 3$\sigma$ of the velocity distribution.    This
suggests against the presence of tidal debris along the line of
sight.   In the absence of evidence to the contrary,  we conclude that
ComBer is in dynamical equilibrium, and therefore current mass
determinations should be robust.

\citet{SimonGeha2007} also report that radial velocity members of
ComBer near the main sequence turn-off lie in a broader color region
than the other ultra-faint dwarfs in their sample, which they
attribute to the effects of multiple stellar populations of different
ages and metallicities.    We have investigated this and find that
this spread is a reflection of larger photometric errors in the
shallower SDSS data.    We will further investigate the stellar
populations of the two objects presented here in a separate paper.

In
contrast to ComBer, UMa\,II shows signs of disequilibrium.
The contours of UMa\,II are elongated, irregular and extend for at least
$1.2^{\circ}$, three times its $r_{\rm half}$.  Only the very
inner core of UMa\,II looks somewhat spheroidal.   Unlike the case of
most other dSphs, UMa\,II's stellar density distribution is not well
matched by any of the commonly used density profiles (King, Plummer or
exponential), and instead it is better matched by two power laws.
Several studies (e.g., \citealt{Johnston1999}; \citealt{Munoz2008};
\citealt{Penarrubia2009}) show that a dwarf satellite initially in a
Plummer or King configuration develops a power law component in the
outer parts as tidal debris is stripped by the Milky Way.  However,
they also show that the inner parts of the satellite retain a
core-like density until the very latest stages of tidal disruption.
The lack of a proper core in the stellar density profile of UMa\,II
may be a indication that this system is in fact in the throes of
destruction.
Another way to assess qualitatively to what extent UMa\,II may have been 
tidally affected, is by comparing the ``break'' in its density profile\footnote{By 
``break'' we mean the inflection point where the inner density profile 
transitions into a power law due 
to the presence of tidal debris.} 
to those of other well-studied disrupting objects such as Pal~5 or the 
Sagittarius (Sgr) dSph. These objects show a clear ``break'' in their 
density distributions around the region where tidal debris is being 
stripped.  In the case of Pal~5 this happens at a surface 
density relative to the central one of $\Sigma_{N,0}/\Sigma_{\rm break}\sim100$ 
whereas for Sgr this value is closer to 200.  If for UMa\,II we take the 
point where the inner power law changes slope as the ``break'' point, 
we obtain $\Sigma_{N,0}/\Sigma_{\rm break}\sim25$, higher than for Pal~5 
and Sgr.  In addition, this break in UMa\,II occurs relatively closer 
to the center than in these two systems, at a break radius 
$r_{b}\approx2\times r_{\rm half}$ compared to $r_{b}\approx4\times r_{\rm half}$ 
for the latter. All these observations further support a tidal scenario for UMa\,II.

\citet{SimonGeha2007} detect a velocity gradient along the major 
axis of UMa\,II, measuring an $8.4\pm1.4$\,km\,s$^{-1}$\ velocity difference
between in eastern and western halves of this object-- in the same
direction as the elongation seen in our deep photometry.   They also
point out that the UMa\,II's velocity dispersion of $6.7\pm1.4$ km
s$^{-1}$ is an outlier in the observed trend of lower velocity
dispersion with decreasing luminosity followed by other Galactic dwarf
galaxies (see their Figure 10a).  Given UMa\,II's absolute magnitude
of $M_{V}=-3.93$, a value of 3--4 km s$^{-1}$ would be more in line
with the observed trend.  One possibility to explain this observation
is that the velocity dispersion of UMa\,II has been inflated by
Galactic tides.  We note however that these observations are limited
to a small region inside $r_{\rm half}$ of the system.  
Despite the
observational evidence discussed above in favor of the tidal stripping
of UMa\,II, our current data set does not allow us to conclude that
this object is completely unbound or out of dynamical equilibrium.
Kinematical data in outer regions are required to determine more
precisely the nature of UMa\,II and to further explore its possible
association with the Orphan Stream.

In the context of the galaxy versus cluster issue raised by
\citet{Gilmore2007,Gilmore2008}, our findings (or lack thereof) imply 
that ComBer is solidly situated in the gap (in $M_{V}$ versus 
$r_{\rm half}$ space) between star clusters and dwarf galaxies, and 
likely cannot be explained away as an evaporating cluster or dissolving 
dwarf galaxy.  Metallicity measurements support this scenario.
While our photometric work presented here show that the stellar content
of ComBer (and UMa\,II) is consistent with being dominated by a very 
metal poor population of [Fe/H]$\sim-2.3$ (see Figs. 3 and 4), a
metallicity unusually low even for the most metal poor globular clusters, 
detailed spectroscopic studies reveal even more extreme numbers.
\citet{Kirby2008} report a mean metallicity for both ComBer and UMa\,II 
of [Fe/H]$\sim-2.5$ which falls squarely on the luminosity-metallicity
relationship found for galaxies.  \citet{Frebel2009} report high
resolution abundances for three stars in ComBer with metallicities
as low as [Fe/H]$\sim-2.9$, which is more metal poor than any of the 
Galactic globular clusters.  In addition, both studies show a metallicity 
spread of 0.6\,dex in this dSph, typical of a dwarf galaxies.
The same argument can be invoked to argue that UMa\,II is a disrupting 
dwarf galaxy as opposed to a dissolving star cluster. 

\section{Conclusions}

We have carried out a deep, wide-field photometric survey of the Coma Berenices and 
Ursa Major II dwarf spheroidal galaxies using the MegaCam Imager on CFHT, reaching 
down to $r\sim25$ mag, more than three magnitudes below the main sequence turn-offs of these Galactic satellites.
This increases roughly by an order of magnitude, with respect to the original SDSS photometry, 
the number of stars that belong to the respective galaxies and that are available determination of 
their structural properties and for morphological studies.

Our results can be summarized as follows:

1. We used a maximum likelihood analysis similar to the one used by \citet{Martin2008}
and \citet{Sand2009} to calculate structural parameters for three different
density profiles: King, Plummer and exponential.
We find characteristic sizes of $r_{\rm half}=74\pm4$\,pc ($5.8\pm0.3\arcmin$) and $123\pm3$\,pc 
($14.1\pm0.3\arcmin$) for ComBer and UMa\,II respectively (from the exponential profile).
Our results provide much tighter constraints on these structural parameters than possible
with previous datasets, but are consistent with earlier determinations using SDSS
photometry.

2. We have re-calculated the total luminosities for both systems and find,
for ComBer $M_{V}=-3.8\pm0.6$ and for UMa\,II $M_{V}=-3.9\pm0.5$ (for a choice of 
Salpeter IMF), in very good agreement with previous results, confirming that ComBer and 
UMa\,II are among the faintest of the known dwarf satellites of the Milky Way.
We have also used a \citet{Chabrier2001} IMF but the results remain virtually unchanged.

3. We have found that ComBer shows a fairly regular morphology with no clear detection 
of potential stripped material down to a surface brightness of $32.4$ mag arcsec$^{-2}$.
Additionally, its number density profile is reasonably well matched by a choice of
either King, Plummer or exponential profile.
We thus conclude that ComBer is likely a stable dwarf galaxy which would
make it one of the most dark matter dominated of the dSph systems.

4. We have also studied the morphology of UMa\,II and find that, unlike ComBer, 
it shows signs of being significantly disrupted.
UMa\,II is larger than previously determined, extending at least
$\sim700$\,pc ($1.2^\circ$ on the sky) and it is also quite elongated.
Its density profile and overall shape resemble a 
structure possibly in the 
latest stages of tidal destruction. 
Furthermore, its number density profile is not well matched by neither of the 
three profiles we tried and it is much better described by two power laws,
further supporting a tidal scenario.

5. The overall 2D surface density distributions of both systems are not
affected by shot-noise and are therefore robust. We find no evidence for isolated 
tidal debris beyond the main bodies of ComBer and UMa\,II to our surface brightness limits of $32.4$ and $34.8$ mag arcsec$^{-2}$, respectively. 

Deep, wide-field imaging of the recently discovered ultra-faint galaxies currently
lags behind spectroscopic observations of these objects. We show in this paper
that high quality, deep photometry is an equally important tool in studying the
dynamical state of the ultra-faint dwarfs. These data can also be used to constrain
the star formation histories of ComBer and UMa\,II which we will explore
in a future contribution.

We acknowledge David Sand, Josh Simon and Gail Gutowski for useful
discussions and Peter Stetson for graciously providing copies of DAOPHOT and ALLFRAME.
M.G and B.W. acknowledge support from the National Science Foundation under
award number AST-0908752.

\begin{deluxetable}{lcccc}
\tablewidth{0pt.}
\tablecaption{Structural Parameters for both ComBer and UMa\,II}
\tablehead{
\colhead{Parameter} & \colhead{Mean} & \colhead{Uncertainty} & \colhead{Mean} & \colhead{Uncertainty}\\
& \colhead{Coma Berenices} & & \colhead{Ursa Major II} 
}
\startdata
$\alpha_{0,exp}$ (h~m~s)&12:26:59.00&$\pm13''$ & 08:51:29.86 &$\pm27''$ \\
$\delta_{0,exp}$ (d~m~s)&+23:54:27.2&$\pm8''$ & +63:07:59.2 &$\pm7''$ \\
$r_{h,exp}$ (arcmin) &5.8&$\pm0.3$ & 14.1 & $\pm0.3$ \\
$r_{h,exp}$ (pc) & 74 & $\pm4$ & 123 & $\pm3$ \\
$r_{h,\rm P}$ (arcmin) &5.9&$\pm0.3$ & 13.9 & $\pm0.3$ \\
$r_{h,\rm P}$ (pc) &76&$\pm4$ & 122 & $\pm3$\\
$\epsilon_{exp}$&0.36&$\pm0.04$ & 0.50 & $\pm0.02$ \\
$\theta_{exp}$ (degrees)&$-$67.0&$\pm3.6$ & $-$74.8 & $\pm1.7$ \\
$N_{*,exp}$ &735&$\pm22$ & 1335 & $\pm35$ \\
\hline
$r_c$ (arcmin)&4.2&$\pm0.5$ & 10.8 & $\pm0.8$ \\
$r_c$ (pc)&54&$\pm6$ & 94 & $\pm7$ \\
$r_t$ (arcmin)&27.9&$\pm3.4$ & 64.2 & $\pm2.8$\\
$r_t$ (pc)&355&$\pm43$ & 560 & $\pm24$\\
\hline
$M_{V}$\tablenotemark{a} (Salpeter)& $-3.8$ & $\pm0.6$ & $-3.9$ & $\pm0.5$\\
$M_{V}$\tablenotemark{a} (Chabrier)& $-3.8$ & $\pm0.6$ & $-4.0$ & $\pm0.6$\\
\enddata \tablenotetext{a}{Using a distance of 44 and 30 kpc for ComBer and UMa\,II respectively, from \citet{Martin2008}}
\end{deluxetable}

\clearpage

\begin{figure}
\plotone{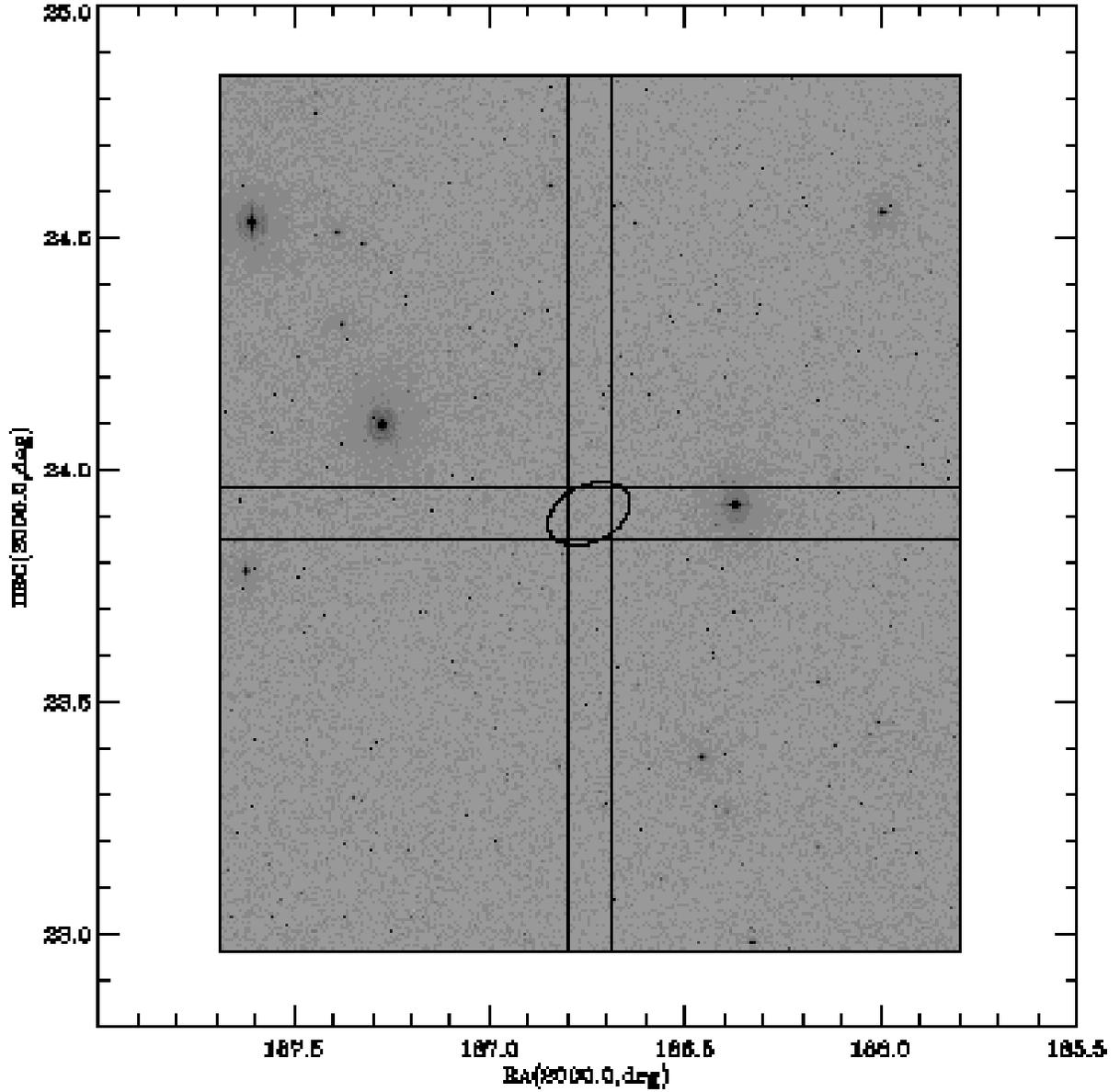}
\caption{Schematic view of our photometric coverage for ComBer. We observed four different 
fields with MegaCam covering roughly $2\times2$ deg$^{2}$ centered on the dSph galaxy. Here, 
the ellipse represents the half-light radius derived by \citet{Martin2008}. A similar observing 
pattern was used for UMa\,II but for a total area of $1.4\times2$ deg$^{2}$.
}
\end{figure}

\begin{figure}
\plotone{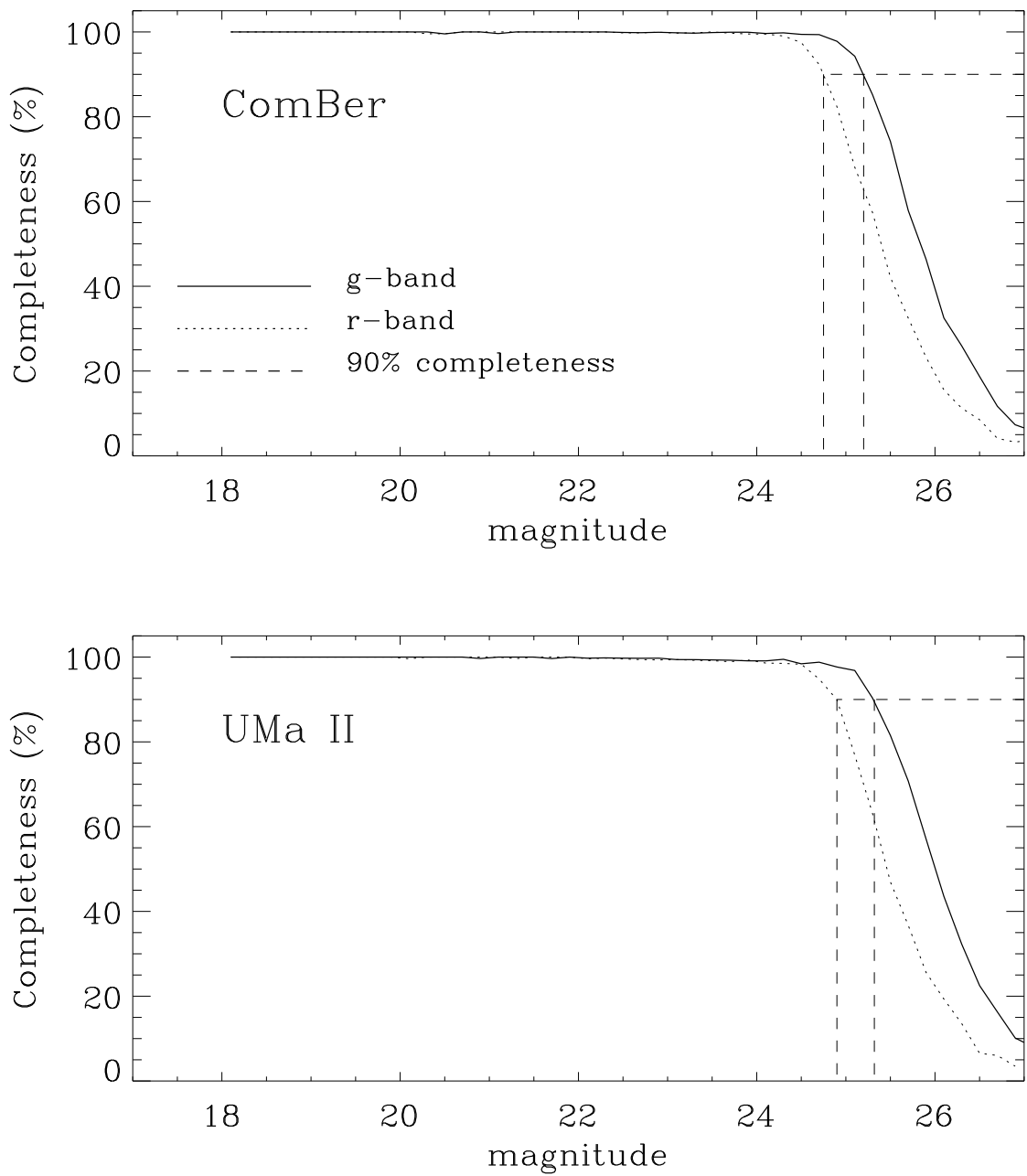}
\caption{Completeness levels as a function of magnitude for our shallowest field
in ComBer (upper panel) and in UMa\,II (lower panel). Solid and dotted lines 
represent the completeness levels as a function of $g$ and $r$ magnitude respectively.
The dashed lines mark the 90\% completeness levels, corresponding to $g=25.2$ and $r=24.75$ 
in the case of ComBer and $g=25.4$ and $r=24.9$ for UMa\,II (See text for further discussion).
}
\end{figure}

\begin{figure}
\plotone{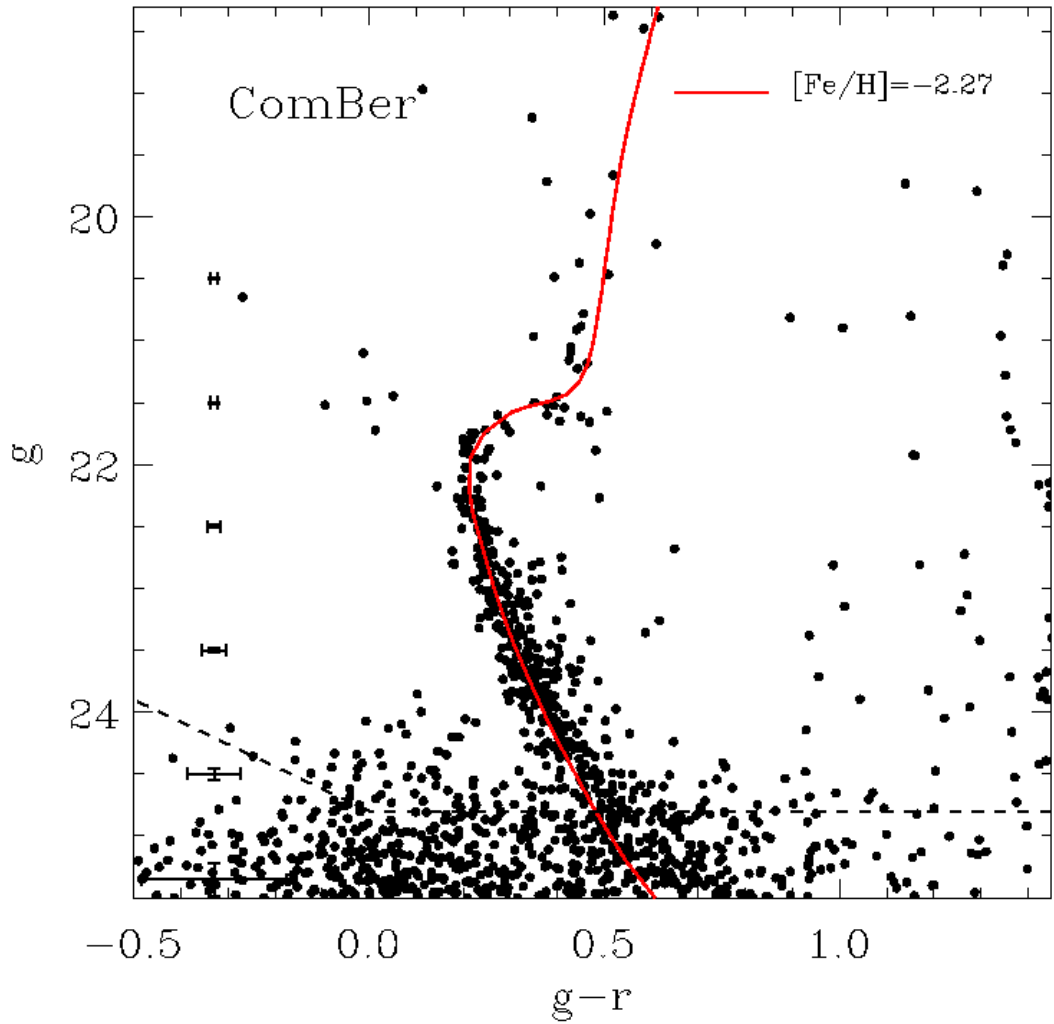}
\caption{Color Magnitude Diagram for the inner region ($r<6\arcmin$) of ComBer.
The dashed lines mark the 90\% completeness level after $\chi$ and $sharp$ cuts
have been applied to remove non-stellar objects. As it can be seen, our CFHT photometry
reaches at least three magnitudes below the main sequence-turn-off of ComBer.
We have complemented our photometry with SDSS data for $g>20$.
The error bars to the left were determined from the artificial star tests and  
represent the standard deviation of a Gaussian function fitted to the error distribution
as a function of magnitude.
A theoretical isochrone for a 13 Gyr old, [Fe/H]=-2.27 population is shown with a solid red line
(from \citealt{Girardi2004}).
}
\end{figure}

\begin{figure}
\plotone{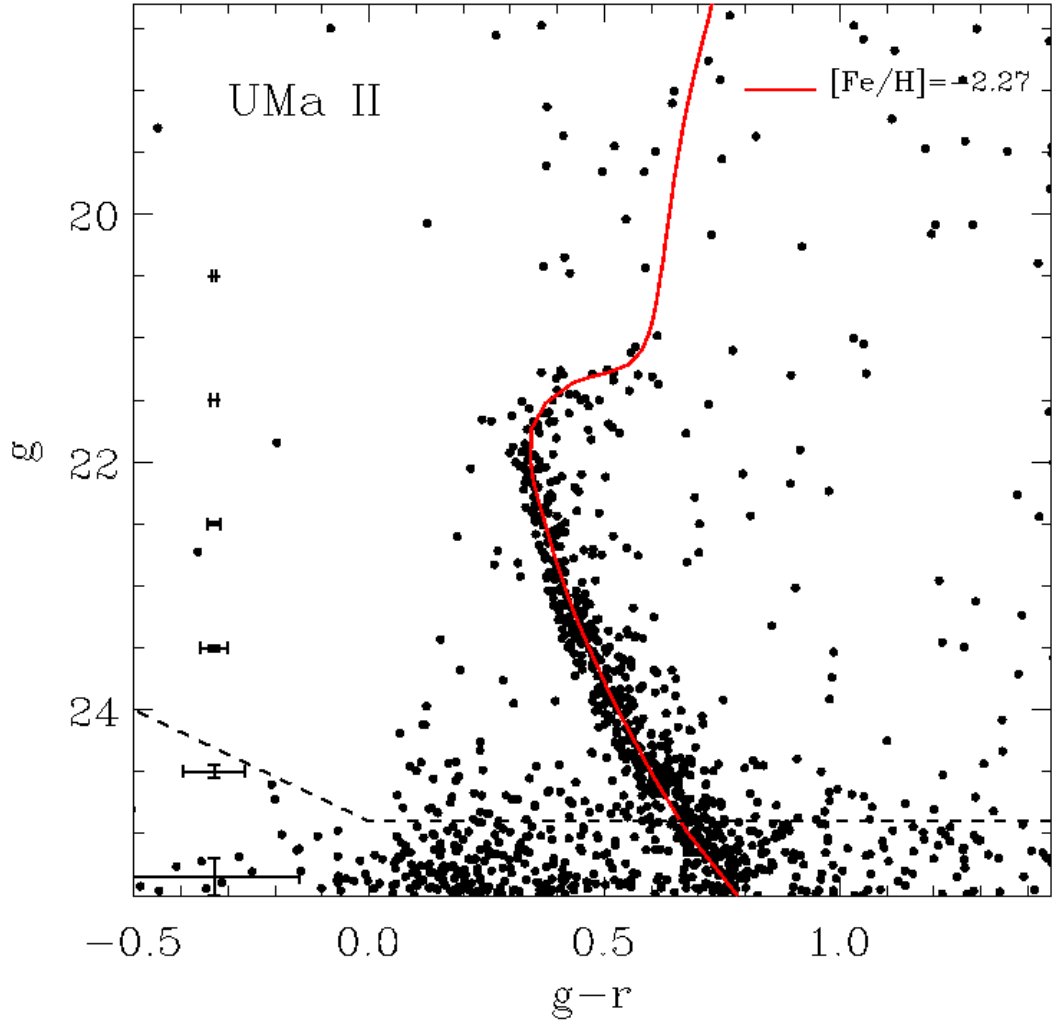}
\caption{Similar to Figure 3 but for UMa\,II.
}
\end{figure}

\begin{figure}
\plotone{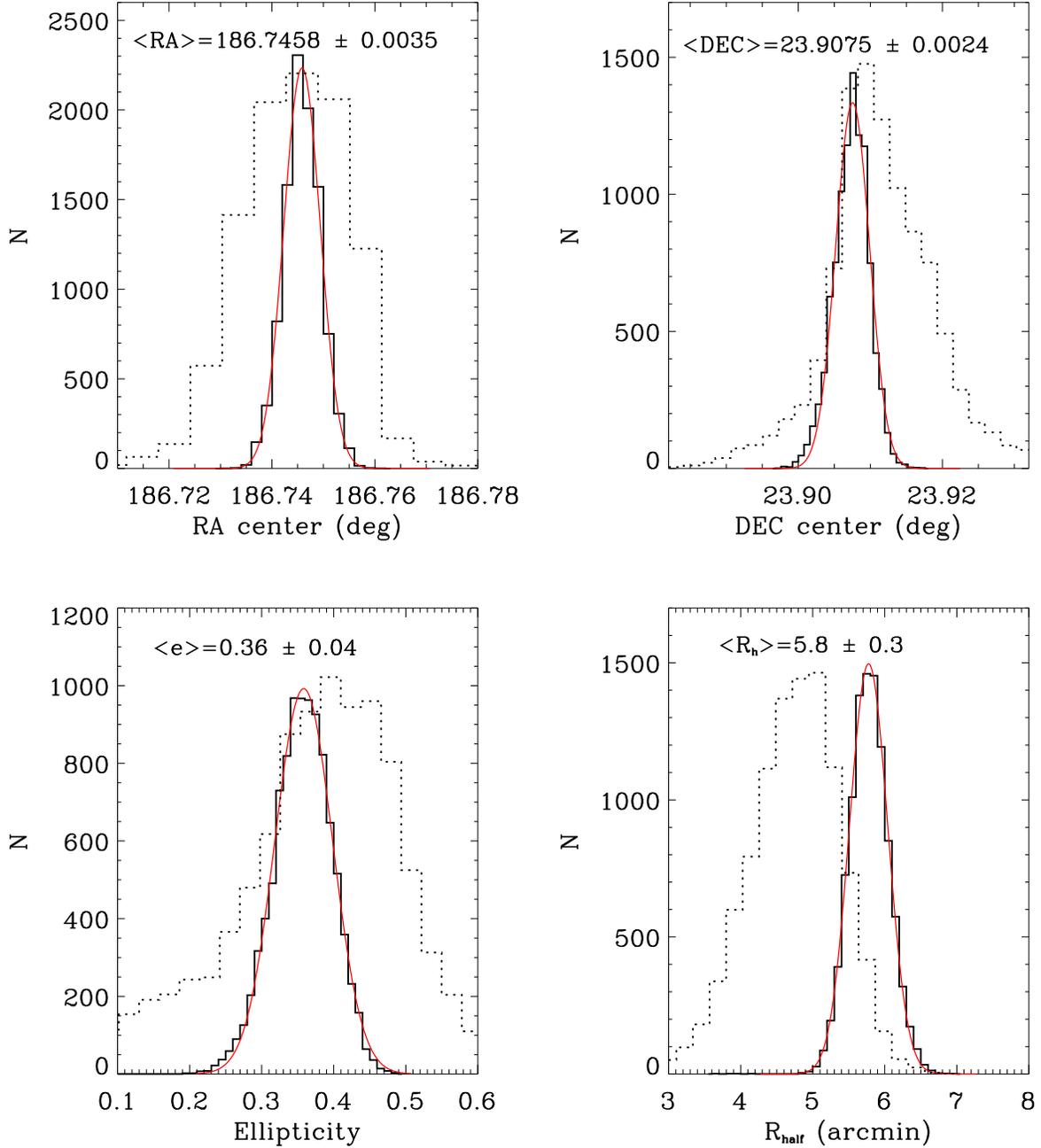}
\caption{Comparison between structural parameters for ComBer derived by applying our ML method
to both the CFHT and SDSS photometry. Solid histograms represent all $10,000$
bootstrap realizations for the CFHT data while dotted histograms show the same but
for SDSS data. The solid red curves show the best-fitted Gaussians to the
CFHT histograms.
}
\end{figure}

\begin{figure}
\plotone{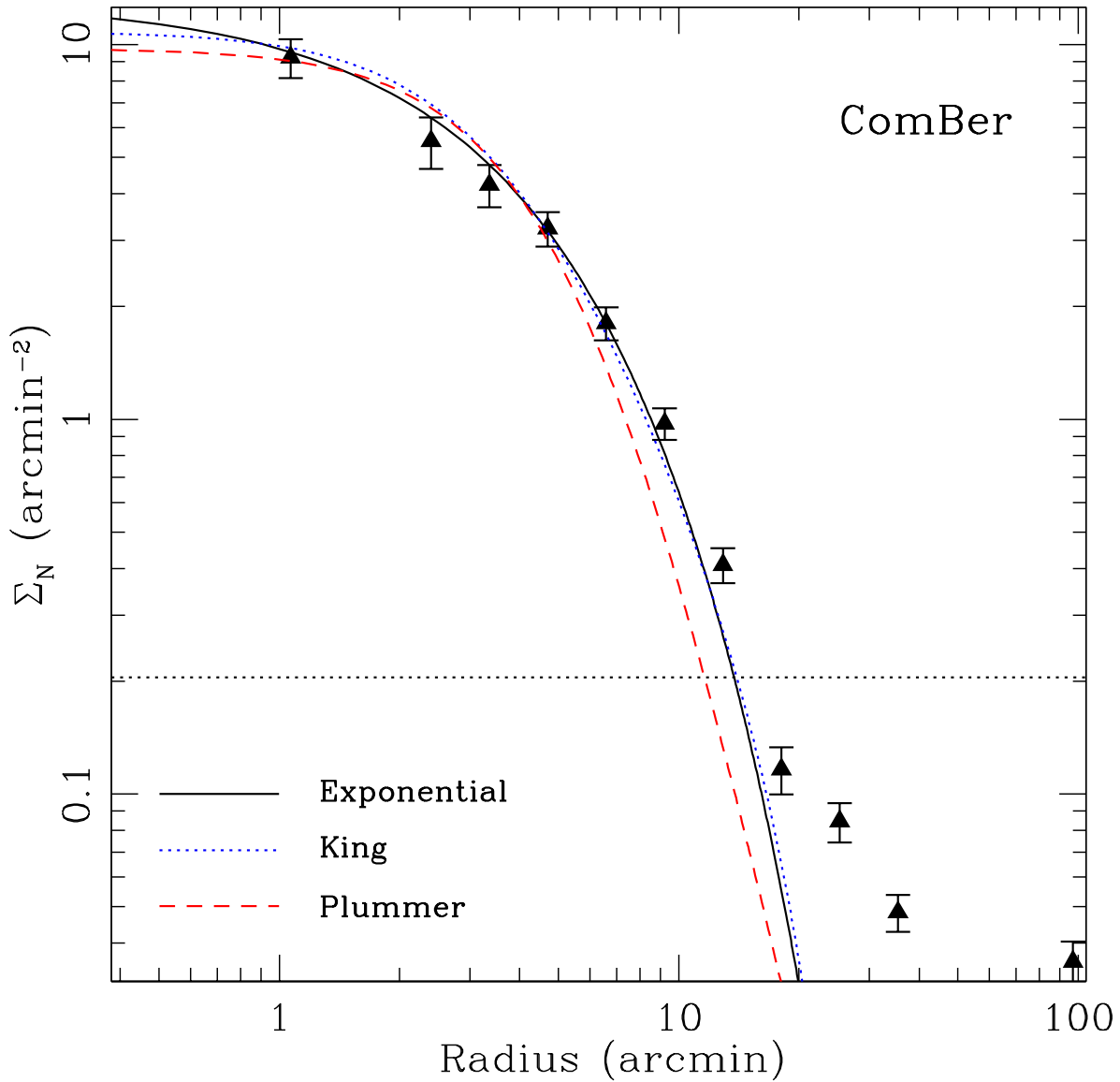}
\caption{Background-substracted number density profile for ComBer. The densities are calculated in
elliptical annuli using the derived structural properties.
Error bars were derived assuming Poisson statistics. 
The horizontal dotted line shows the level of the subtracted background.
Solid (black), dotted (blue) and dashed (red) lines represent the best exponential, 
King and Plummer profiles respectively.
}
\end{figure}

\begin{figure}
\plotone{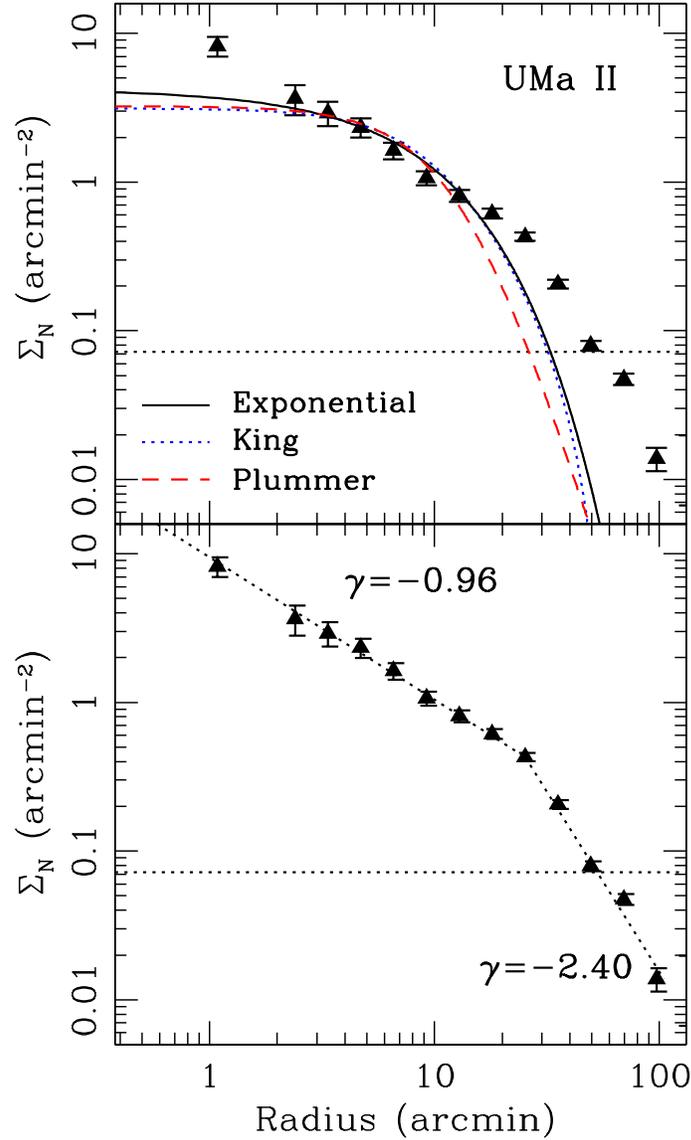}
\caption{{\it Upper panel}: similar to Figure 6, but for UMa\,II. Note that neither profile represents 
a good fit to the observed data points. {\it Lower panel:} background subtracted number density profile 
of UMa\,II where the data points have been fitted with two power laws. These lines represent a much 
better visual fit to the data points than the density profiles used to derive UMa\,II' structural parameters.
}
\end{figure}

\begin{figure}
\plotone{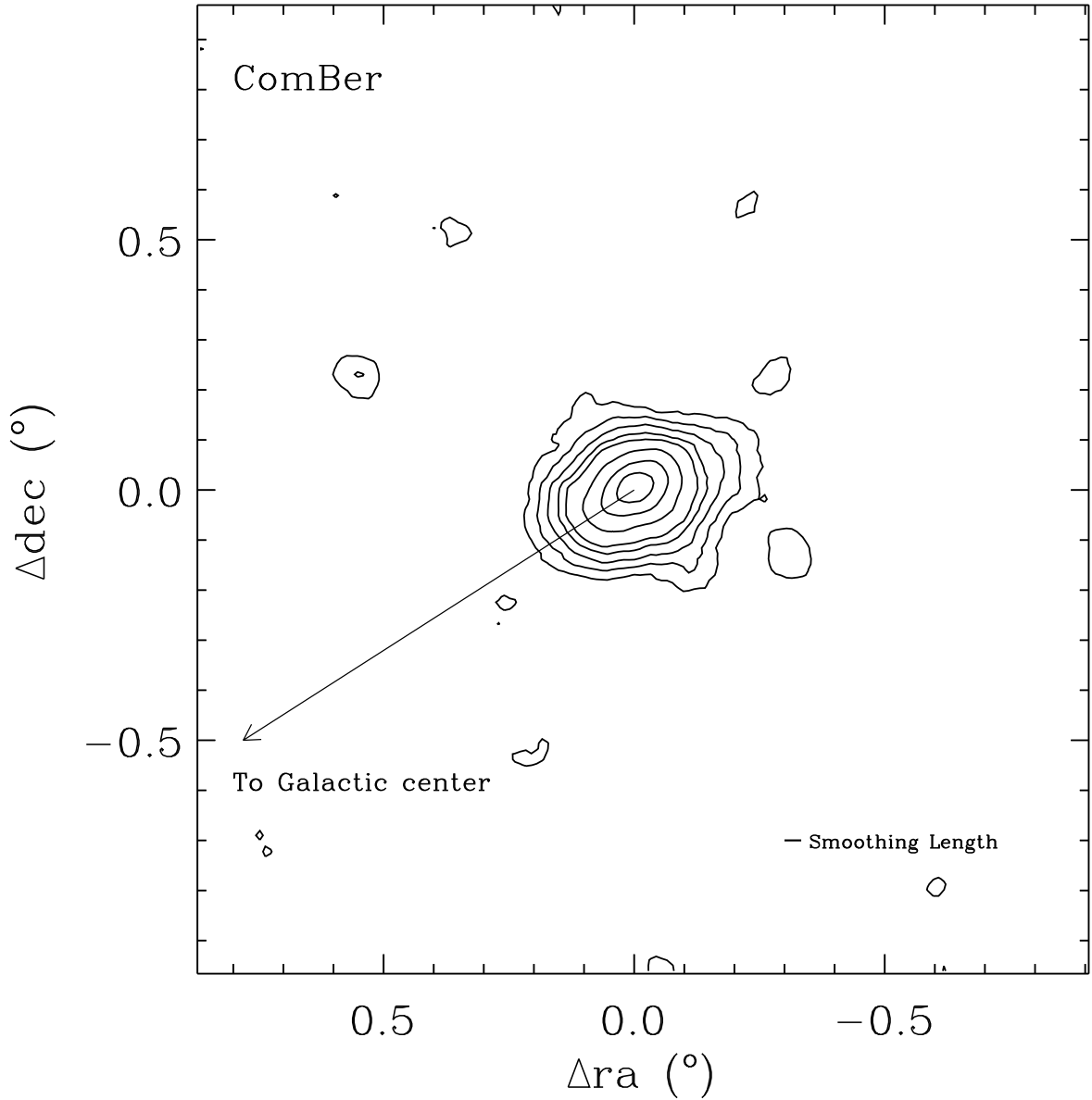}
\caption{Isodensity contour map for ComBer. The contours represent 
$3$, $6$, $10$, $15$, $20$, $35$, $55$ and $75\sigma$ above 
the mean density measured away from the main body of ComBer.
The solid line shows the direction toward the Galactic center.
The smoothing scale length of $2\arcmin$ is also indicated in the figure.
}
\end{figure}

\begin{figure}
\plotone{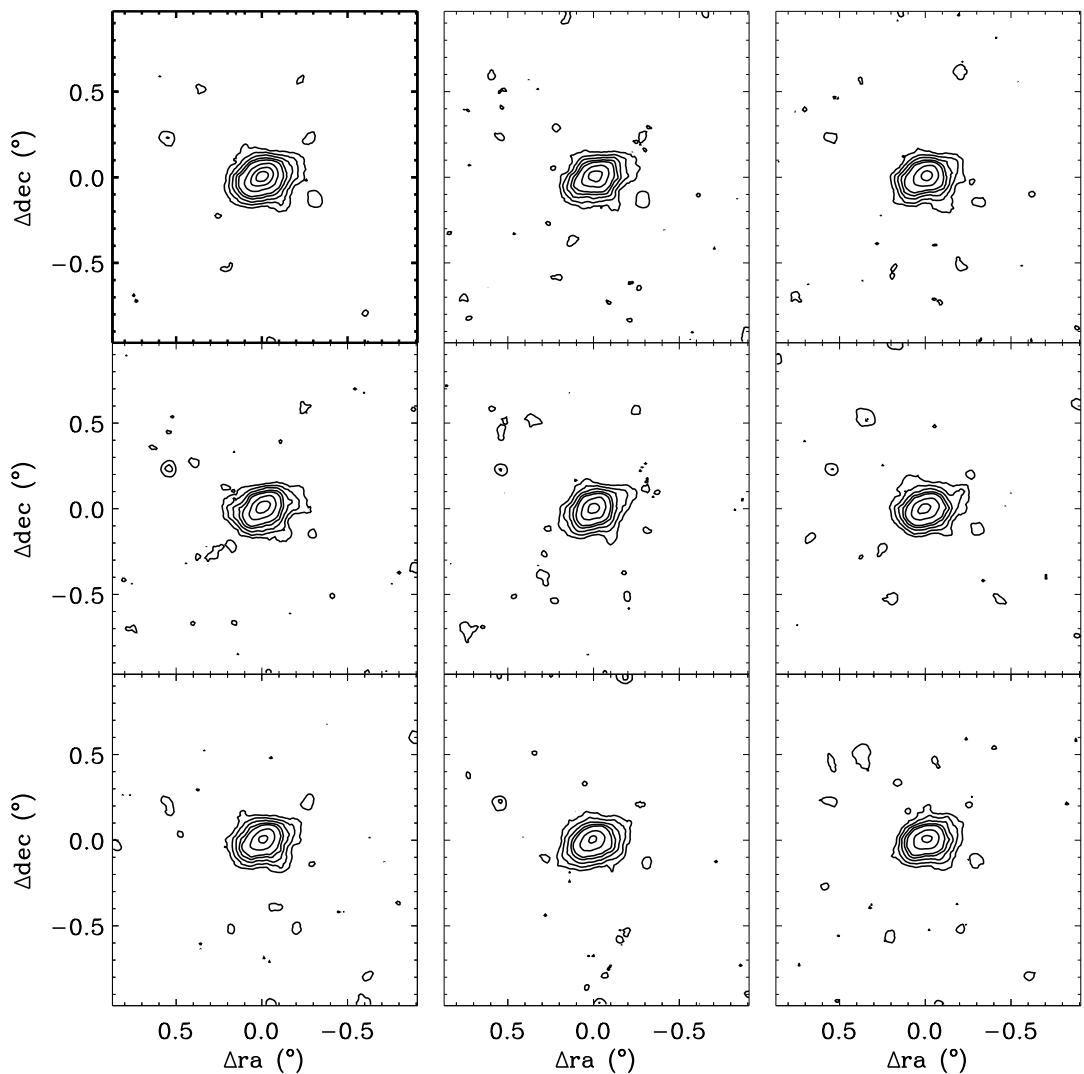}
\caption{Isodensity contour maps for eight random bootstrap realizations
of our photometric data for ComBer. The upper left panel (thicker axis) 
shows the contours for the actual data. As in Figure 8, we show
$3$, $6$, $10$, $15$, $20$, $35$, $55$ and $75\sigma$ above 
the mean density. While the lowest $3\sigma$ contours are not statistically
significant, all more significant contours are well stablished with these data.
} 
\end{figure}

\begin{figure}
\plotone{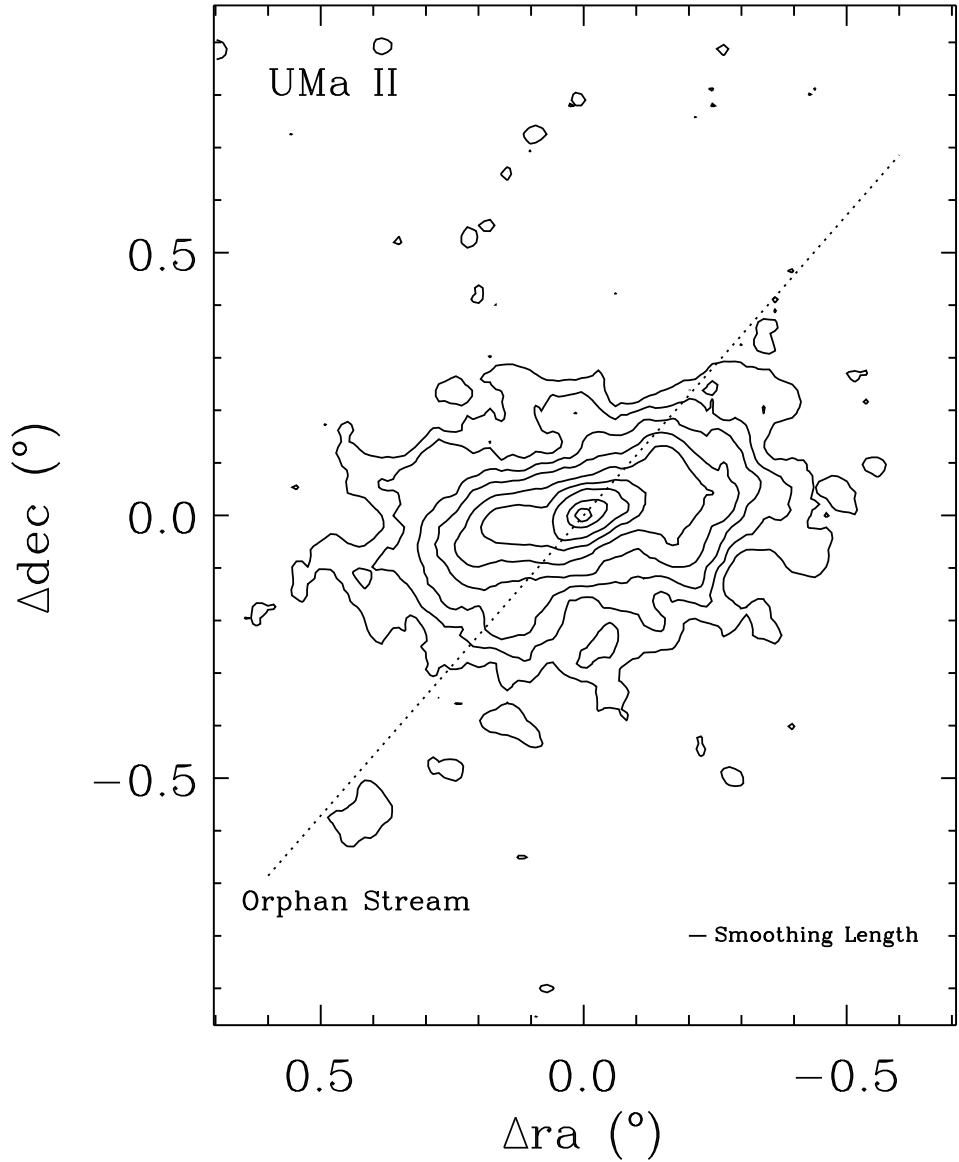}
\caption{Isodensity contour map for UMa\,II.
The contours in this case represent $3$, $6$, $10$, $16$, $26$, $38$, $55$ 
and $70\sigma$ above the density measured in regions away from UMa\,II.
The solid line marks the direction of the orbit derived by \citet{Fellhauer2007} assuming
that UMa\,II is associated with the Orphan Stream. As in Figure 8, the smoothing scale
length is shown for reference.
}
\end{figure}

\begin{figure}
\plotone{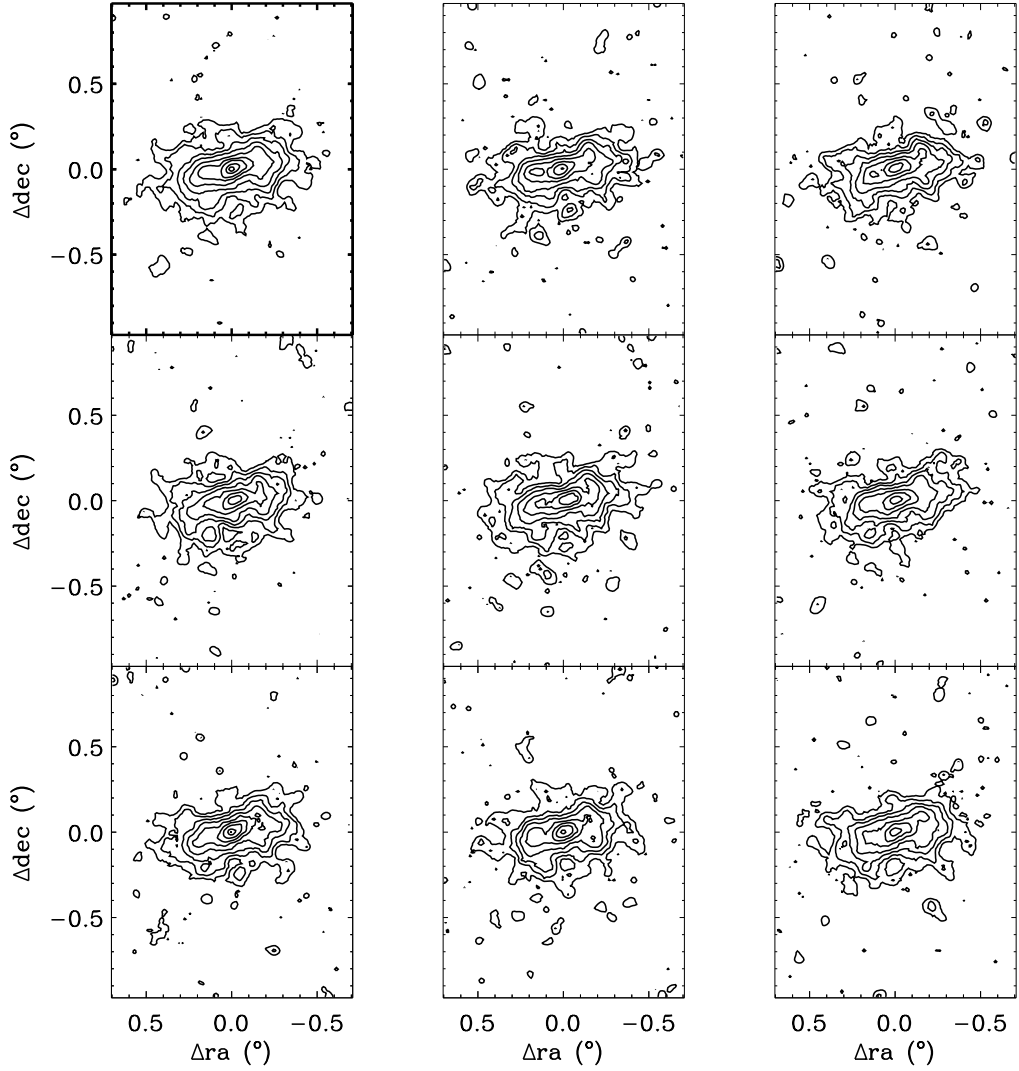}
\caption{Similar to Figure 9 but for UMa\,II. The contours shown are similar
to Figure 10, i. e., 
$3$, $6$, $10$, $16$, $26$, $38$, $55$ and $70\sigma$ above the mean density.
}
\end{figure}

\end{document}